\documentstyle[aps,preprint]{revtex}

\def\l{\lambda}
\def\o{\omega}
\def\s{\sigma}
\def\D{\Delta}
\def\be{\begin{eqnarray}}
\def\ee{\end{eqnarray}}
\def\nn{\nonumber}

\def\e{\epsilon}
\def\h{{1\over 2}}

\def\.{\cdot}
\def\({\left(}
\def\[{\left[}
\def\){\right)}
\def\]{\right]}
\def\C{\overline{C}}
\def\U{\overline{U}}

\def\eq#1{(\ref{#1})}
\def\labels#1{\label{#1}}
\begin{document}

\title{Time Ordering, Energy Ordering, and Factorization}        
\author{C.S. Lam}        
\address{Department of Physics, McGill University\\
3600 University St., Montreal, QC, Canada H3A 2T8}
\maketitle

\begin{abstract}
Relations between integrals of time-ordered product of operators,
and their representation in terms of energy-ordered products
are studied. Both can be decomposed into irreducible factors
and these relations are discussed as well. The energy-ordered
representation was invented to separate various infrared contributions
in gauge theories. It is shown that the irreducible time-ordered
expressions can be used to accomplish the same purpose. Besides, 
it has the added advantage of being factorizable.
\end{abstract}

\section{Introduction}
Integrals of time-ordered products
are often encountered in physics. For example, the $n$th order
high-energy tree amplitude $U_n$ in the presence of an interaction
$H(t)$, is given by the integral of the time-ordered product of
$n$ such operators. In the energy representation, vertices of
the scattering amplitude are given by the Fourier transform $h(\o)$ of
$H(t)$, and energy denominators emerge from the time-ordered integrations.

It is known that $U_n$ can be 
decomposed into sums of products of {\it irreducible amplitudes}
$C_{m_i}$, with $\sum_im_i=n$ \cite{LL1JMP}. 
The irreducible amplitudes $C_m$ are identical
to the scattering amplitudes $U_m$, except the time-ordered products
of $H(t_i)$'s  are replaced by their time-ordered
nested commutators. In the energy representation, ordinary product
of $h(\o_i)$'s turn into their nested commutators.

This decomposition formula embraces a {\it factorization} property
which turns out to be very useful. A review of its applications can be found in
\cite{LRECOIL}. Its basic properties will be reviewed in Sec.~2.

On the other hand, 
it is also known \cite{CC} that the time-ordered-product expression for
$U_n$ can be turned into an
energy-ordered products containing the 
nested commutators of $h(\o_i)$'s. This will be reviewed in Sec.~3.
The energy-ordered expressions are much more complicated than the original
time-ordered expressions, but they are useful in sorting out the
leading and the subleading 
contributions in the infrared regime of QCD. This will be
briefly discussed in Sec.~6.

Since nested commutators appear in the decomposition formula, and
in the energy-ordered expression, one might wonder whether
the two formulas are essentially  the same
\footnote{This investigation
is partly motivated by questions raised by Prof. M. Ciafaloni and
Prof. S. Catani, regarding the
distinction between these two approaches. I am grateful to them for
their questions and the ensuring discussions.}. 
In spite of the superficial
similarity, these two are actually quite different.
The energy-ordered formula turns time-ordering into energy-ordering,
but the whole expression remains {\it ordered},
 with `propagators' {\it linking} the
whole amplitude. On the other hand, the decomposition formula seeks
to {\it factorizable} the integral of any {\it ordered product} into
disjoint irreducible factors. It can be
applied to time-ordered expressions,
and it can also be applied to energy-ordered formulas,
as will be discussed in Sec.~4. There are irreducible factors
$C_n$ for the time-ordered amplitudes, and separately irreducible
factors $\C_N$ for energy-ordered amplitudes. The relations between
the time-ordered quantities and the energy-ordered quantities
will be discussed in Sec.~5. At low orders these relations can be
directly verified, as done in the Appendix, but at high orders
the algebra is so complicated that it is impractical to 
obtain these relations by brute force.

As mentioned above, 
the energy-ordering formula is useful in sorting out the leading
from the subleading terms in the infrared region of QCD \cite{CC}. 
It will be shown in Sec.~6 that the decomposition formula
applied {\it directly}
to time-ordered products can be used for that purpose as well.
Moreover, the factorization property inherent in the decomposition 
formula gives
this approach an added advantage that  will be discussed in Sec.~6.

\section{Decomposition Formula}
Let $H(t)$ be an operator-valued function of $t$, and 
\be
U_n(TT')&=&{1\over n!}\int_{T'}^Tdt_1dt_2\cdots dt_n\(H(t_1)H(t_2)\cdots H(t_n)\)_+\nn\\
&=&\int_{R_n(TT')}dt_1dt_2\cdots dt_nH(t_1)H(t_2)\cdots H(t_n)
\labels{top}\ee
be the integral of its time-ordered product, denoted by $(\cdots)_+$.
The integration region $R_n(TT')$ is the hyper-triangular region 
defined by
$\{T\ge t_1\ge t_2\ge\cdots\ge t_n\ge T'\}$.
The time-ordered exponential is related to them by
\be
U(TT')\equiv\(\exp\[\int_{T'}^TH(t)dt\]\)_+=1+\sum_{n=1}^\infty U_n(TT').
\labels{toe}\ee
Each $U_n$ can be decomposed into sums of products of {\it irreducible
 components} $C_m$ via the formula \cite{LL1JMP}
\be
U_n&=&\sum_{\{m\}}\xi(m_1m_2\cdots m_k)C_{m_1}C_{m_2}\cdots C_{m_k},
\labels{dec}\\
\xi(m)&\equiv&\xi(m_1m_2\cdots m_k)=\prod_{i=1}^k\(\sum_{j=i}^km_j\)^{-1},
\labels{xi}\ee
where the sum in \eq{dec}
is taken over all partitions $(m)=(m_1m_2\cdots m_k)$ of the number
$n$, so that
$m_i>0$ and $\sum_{i=1}^km_i=n$. The irreducible components $C_m$
are defined similar to $U_m$, but with the time-ordered product
replaced by time-ordered nested commutators:
\be
C_n(TT')&=&{1\over n!}\int_{T'}^Tdt_1dt_2\cdots dt_n
\(H[t_1t_2\cdots t_n]\)_+\nn\\
&=&\int_{R_n(TT')}dt_1dt_2\cdots dt_n
H[t_1t_2\cdots t_n],
\labels{cn}\\
H[t_1t_2\cdots t_n]&\equiv&
[H(t_1),[
H(t_2),[\cdots,[H(t_{n-1}), H(t_n)]\cdots]]],\nn\\
H[t_1]&\equiv&H(t_1).
\labels{nestedcom}\ee
Note that \eq{dec} is combinatorial in nature, so 
the decomposition is valid 
whether the parameters $t$ is `time', or `energy', or anything 
else.

As an illustration, I list below the explicit decomposition formula for
the first three $U_n$'s:
\be
U_1&=&C_1,\nn\\
U_2&=&{1\over 2}C_1^2+{1\over 2}C_2,\nn\\
U_3&=&{1\over 6}C_1^3+{1\over 3}C_2C_1+{1\over 6}C_1C_2+{1\over 3}C_3.\labels{u3c}\ee

The $n$ operators used
in the time-ordered product in \eq{top} are identical.
If they are all different, a decomposition formula similar to \eq{dec}
still exists. This will be discussed further in Sec.~6.

Of particular interest is
the `high-energy off-shell tree amplitude' $U_n(0-\infty)$. 
It may be expressed
in terms of the Fourier transform $h(\o)$ of $H(t)$,
\be
H(t)=\int_{-\infty}^\infty h(\o)e^{-i\o t}d\o,
\labels{fourier}\ee
to be
\be
U_n(0-\infty)&=&{i^n\over n!}\int_{-\infty}^\infty h(\o_1)h(\o_2)\cdots
h(\o_n)\prod_{i=1}^n{d\o_i\over\Omega_i+i\e},\labels{vert}\\
\Omega_i&=&\sum_{j=i}^n\o_j,
\labels{prop}\ee
where $h(\o_i)$ are the vertices, and
the propagators $1/(\Omega_i+i\e)$ are obtained from \eq{top} and 
\eq{fourier} by carrying out the time integrations explicitly.
The corresponding off-shell irreducible
amplitude $C_n$ is
\be
C_n(0-\infty)&=&{i^n\over n!}\int_{-\infty}^\infty 
h[\o_1\o_2\cdots
\o_n]\prod_{i=1}^n{d\o_i\over\Omega_i+i\e},
\labels{cne}\ee
where $h[\o_1\o_2\cdots\o_n]$ is the nested commutator of $n$
$h(\o_i)$, defined similar to \eq{nestedcom}.

\section{Energy-Ordering Formula}
The time-ordered exponential $U(0-\infty)$ may be converted into an
energy-ordered exponential in the following way \cite{CC}.
First, replace $H(t)$ everywhere by
\be
H_E(t)=\int_{-\infty}^Eh(\o)e^{-i\o t}d\o.\labels{he}\ee
The corresponding time-ordered exponential \eq{toe} will be denoted
by $U_E(0-\infty)$. It satisfies the following
 differential equation in $E$,
\be
{dU_E(0-\infty)\over dE}&=&\int_{-\infty}^0dt\ 
U_E(0t)h(E)e^{-iE t}U_E(t-\infty)
\equiv \D(E)U_E(0-\infty),\nn\\
&&\labels{diffeq}\ee
where
\be
\D(E)&\equiv&\int_{-\infty}^0dt\ e^{-iE t}U_E(0t)h(E)
U_E(0t)^{-1}
\equiv\sum_{n=1}^\infty\D_n(E),\labels{deln}\\
\D_n(E)&=&\int_{-\infty}^0 dt_1\int_{-\infty}^{t_1}dt_2\cdots 
\int_{-\infty}^{t_{n-1}}dt_n\ e^{-iE t_n}
\nn\\
&&[H_E(t_1),[H_E(t_2),[\cdots,[H_E(t_{n-1}),h(E)]\cdots
]]].\labels{dne1}\ee
Using \eq{he} and carrying out the time integrations, we get
\be
\D_n(\o_n)&=&i^{n}\int_{-\infty}^{\o_n}
h[\o_1\o_2\cdots\o_n] \(\prod_{i=1}^{n-1}{d\o_i\over\Omega_i+i\e}\)
{1\over\Omega_n+i\e}.
\labels{eoepf}\ee
Please note a deceiving similarity between 
$\int_{-\infty}^\infty \D_n(\o_n)d\o_n$ and $C_n(0-\infty)$.
The two would have been the same if the upper limit of the 
$(n-1)$-dimensional integrations in \eq{eoepf} were $\infty$ instead
of $\o_n$. This turns out to be the important distinction between the two
approaches.

 Integrating the
differential equation \eq{diffeq} and letting $E\to\infty$, 
we finally obtain an energy-ordered
expression for $U(0-\infty)$ to be \cite{CC}
\be
U(0-\infty)&=&
\(\exp\[\int_{-\infty}^\infty
 \Delta(\o)d\o\]\)_+\equiv 1+\sum_{N=1}^\infty\U_N,\labels{ute}\\
\U_N(0-\infty)&=&{1\over N!}
\int_{-\infty}^\infty d\o_1d\o_2\cdots d\o_N \(\D(\o_1)\D(\o_2)\cdots\D(\o_N)\)_+\nn\\
&=&\int_{R_N(\infty-\infty)}d\o_1d\o_2\cdots d\o_N \D(\o_1)\D(\o_2)\cdots\D(\o_N),
\labels{en}\ee
where $R_N(\infty-\infty)$ is the hyper-triangular integration region
$\infty>\o_1\ge \o_2\ge\cdots\ge\o_N>-\infty$.

Before proceeding further let us pause to
compare the time-ordered and the
energy-ordered expressions for $U(0-\infty)$. The time-ordered
expression, given in \eq{toe}, has an exponent linear in $H(t)$, or in
$h(\o)$. The energy-ordered expression, given in \eq{ute}, has an
exponent
containing all orders of $h(\o)$ because of \eq{deln} and \eq{eoepf}.
Similarly, $U_n$ given in \eq{vert} and $C_n$
given in \eq{cne} are of order $n$, but
$\U_N$ in \eq{en} contains all orders from $N$ on. It would be 
convenient to be able to refer to each order separately, so let us
define $\U_{Nn}$, with $n\ge N$, to be the order-$n$ terms of $\U_N$.
Hence
\be
\U_N=\sum_{n=N}^\infty\U_{Nn}.
\labels{unn}\ee

\section{Decomposition of the Energy-Ordering Formula}
The energy-ordered formula in \eq{en} for the amplitude $\U_N$
can be decomposed into irreducible components using the
decomposition formula \eq{dec}:
\be
\U_N&=&\sum_{\{M\}}\xi(M_1M_2\cdots M_k)\C_{M_1}\C_{M_2}\cdots \C_{M_K},
\labels{ubcb}
\ee
where the sum is taken over all partitions
$(M)=(M_1M_2\cdots M_K)$ of the number $N$.
In particular, as in \eq{u3c}, we have
\be
\U_1&=&\C_1,\nn\\
\U_2&=&{1\over 2}\C_1^2+{1\over 2}\C_2,\nn\\
\U_3&=&{1\over 6}\C_1^3+{1\over 3}\C_2\C_1+{1\over 6}\C_1\C_2+
{1\over 3}\C_3.
\labels{u3e}\ee

The irreducible components are now given by
\be
\C_N&=&{i^N\over N!} 
\int_{-\infty}^\infty d\o_1d\o_2\cdots d\o_N 
\(\D[\o_1\o_2\cdots\o_N)]\)_+\nn\\
&=&i^N\int_{R_N(\infty-\infty)}d\o_1d\o_2\cdots d\o_N \D[\o_1\o_2\cdots\o_N]\nn\\
&\equiv&\sum_{n=N}^\infty\C_{Nn},\labels{enc}\ee
where $\D[\o_1\o_2\cdots\o_N]$ is the nested commutator of $N$
$\D(\o_i)$'s, defined similar to \eq{nestedcom}, and $\C_{Nn}$
is the term in $\C_N$ of order $n$.

Unlike \eq{dec} and \eq{u3c}, where each decomposition contains
quantities of a fixed order, the decomposition
\eq{ubcb} and \eq{u3e} each contains quantities
of all orders. By equating quantities of the same order, we obtain
from each equation an infinite number of identities. Up to order 3,
these are:
\be
\U_{1n}&=&\C_{1n},\nn\\
\U_{22}&=&{1\over 2}\C_{11}^2+{1\over 2}\C_{22},\nn\\
\U_{23}&=&{1\over 2}\C_{12}\C_{11}+{1\over 2}\C_{11}\C_{12}+
{1\over 2}\C_{23},\nn\\
\U_{33}&=&{1\over 6}\C_{11}^3+{1\over 3}\C_{22}\C_{11}+
{1\over 6}\C_{11}\C_{22}+
{1\over 3}\C_{33}.
\labels{u3ed}\ee

\section{Relations between Time-Ordered and Energy-Ordered Quantities}
By equating the time-ordered and the energy-ordered expressions for
$U(0-\infty)$ at a given order, relations between the two kinds of quantities can be obtained. To start with,
\be
U_n=\sum_{N=1}^n\U_{Nn}.
\labels{uu}\ee
For the first few orders, we have
\be
U_1&=&\U_{11},\nn\\
U_2&=&\U_{12}+\U_{22},\nn\\
U_3&=&\U_{13}+\U_{23}+\U_{33}.\labels{uue}\ee

By using \eq{dec} and \eq{ubcb}, we can also obtain from \eq{uu} a relation
between the irreducible amplitudes $C_n$ and $\C_{Nn}$. The first few
of these relations are
\be
C_1&=&\C_{11},\nn\\
C_2&=&2\C_{12}+\C_{22},\nn\\
C_3&=&\C_{33}+{3\over 2}\C_{23}+3\C_{13}+{1\over 2}[\C_{11},\C_{12}].
\labels{bub}\ee

In the Appendix, we shall write down the low-order quantities 
in their explicit forms 
 as  integrals of $h(\o)$ and the propagators. We
shall then verify directly that these identities are valid.
The algebra encountered are fairly complicated, showing that these
quantities are really intertwined in a very complicated way.

\section{Infrared Behavior}
Imagine $h(\o)$ to be the vertex emitting ($\o>0$) 
or absorbing ($\o<0$) soft photons 
or gluons of energy $|\o|$. For now let us consider only emissions
so that we may assume $h(\o)=0$ for $\o<0$. As a result, the integration
region of all the $\o$-integrals may be restricted between 0 and $\infty$.
If $h(\o)$ approaches a non-zero operator as $\o\to 0^+$, say of order
$g$, then an infrared
divergence occurs in the scattering amplitude $U_n$ in \eq{vert}. 
If $\lambda$ is the infrared cutoff for the $\o$-integrations, then
the leading infrared divergence of $U_n$ is seen from \eq{vert} and
\eq{prop} to be of order $g^n\log^n\lambda$, 
and that is produced in the strongly ordered region
\be
\o_1\gg\o_2\gg\cdots\gg\o_n\gg\l.\labels{oregion}\ee
The subleading contribution, of order
$g^n\log^m\l$, comes from regions where $m$ $\o_i$'s are strongly ordered
like \eq{oregion}, with the rest of the $\o_i$'s of the same general
magnitude as one of those in the ordered region. 

\subsection{Energy ordering}
The energy-ordering formula discussed in Sec.~3 offers
a systematic way to extract from $U_n$ terms with subleading contributions
\cite{CC}.
The basic observation is that 
 $\D_n(\o_n)$ of \eq{eoepf}  is of order $g^n\log\l$ whatever $n$ is.
That behavior arises because the variable
$\o_n$ in the propagator $1/(\Omega_n+i\e)=1/(\o_n+i\e)$
is the largest (rather than the smallest) variable in \eq{eoepf}. As
a result, the magnitude of a term is controlled by the number of $\D_n$'s
it contains. In particular, $\U_{Nn}$ would be of order $g^n\ln^N\l$.

This fact can be used to analyse the infrared behavior of QCD  \cite{CC}, but we will not go into the details here. 
Instead, we would like 
to discuss an alternative method to extract subleading terms directly 
from the time-ordered expression, with the help of the
decomposition formula.

At first sight this seems to be impossible. 
As mentioned above, the reason why
$\D_n\sim g^n\ln\l$ rather than $g^n\ln^n\l$, is because
$\o_n$ in \eq{eoepf} is larger than any other energy 
variables in the integral. 
As pointed out below \eq{eoepf},
this is precisely a property that $C_n(0-\infty)\equiv C_n$ 
does not share, so it is hard 
to imagine why decomposition into $C_m$'s using \eq{dec} can achieve
the same end. The remedy, it turns out, is to decompose the 
{\it integrand} of $U_n$ rather than the integral itself.

\subsection{Infrared decomposition from time-ordering}
We will continue to assume $h(\o)=0$ for $\o<0$ so that the 
integration region in \eq{vert} becomes the hyper-cube
$W_n=\{\infty>\o_i\ge 0,\ 1\le i\le n\}$.
This integral can be rewritten as an integral 
over the hyper-triangular region  
\be
R_n=\{\infty>\o_1\ge\o_2\cdots\ge\o_n\ge 0\},
\labels{triang}\ee
provided we symmatrize the integrand:
\be
U_n&=&{i^n\over n!}\int_{R_n}d^n\o \ u_n,\nn\\
u_n&=&\sum_{[\s]\in S_n}u[\s],\nn\\
u[\s]&=&h(\o_{\s_1})h(\o_{\s_2})\cdots h(\o_{\s_n})\prod_{i=1}^n
{1\over \Omega^\s_i+i\e},\nn\\
\Omega^\s_i&=&\sum_{j=i}^n\o_{\s_j}.
\labels{unint}\ee
In these formulas, 
$[\s]\equiv[\s_1\s_2\cdots\s_n]$ is a permutation of $[12\cdots n]$,
and the sum is taken over the symmetric group $S_n$ of all such permutations.

The integrand $u_n$ 
can be decomposed into irreducible amplitudes $c[\s']$
in a way similar to \eq{dec} \cite{LL1JMP,LRECOIL},
\be
u_n=
\sum_{[\s]\in S_n}c[\s]_P,\labels{cp}\ee
where $c[\s]_P$ indicates a product of irreducible factors $c[\s']$
similar to the right-hand side of \eq{dec}.
Given a sequence $[\s]=[\s_1\s_2\cdots\s_n]$, 
we partition it by inserting
vertical bars as follows: a bar is inserted after a number
$\s_i$ if and only if it is smaller than every number to its right.
For example, if $[\s]=[12345]$, then 
$[\s]_P=[1|2|3|4|5]$. If $[\s]=[52134]$, then $[\s]_P=[521|3|4]$, and if
$[\s]=[14352]$, then  $[\s]_P=[1|43|52]$. Such partitions divide
the original sequence $[\s]$ into subsequences, separated by the vertical
bars. For the infrared behavior to be discussed later, it is important
to note that the last (rightmost) number of every subsequence is
the smallest number of that subsequence.  Now
$c[\s]_P$ is simply the product of 
the irreducible factors $c[\s']$, one for each subsequence. 
For example, $c[12345]_P=
c[1]c[2]c[3]c[4]c[5]$, $c[52134]_P=c[521]c[3]c[4]$, and 
$c[14352]_P=c[1]c[43]c[52]$. The {\it irreducible amplitude} $c[\s']=
c[\s'_1\s'_2\cdots \s'_m]$ is given  by
\be
c[\s']&=&c[\s'_1\s'_2\cdots\s'_n]=
h[\s']\prod_{i=1}^m
{1\over\Omega^{\s'}_i+i\e},\nn\\
h[\s']&\equiv &h[\s'_1\s'_2\cdots\s'_m]\equiv[h(\s'_1),[h(\s'_2),
[\cdots,[h(\s'_{m-1}),h(\s'_m)]\cdots]]],\nn\\
h(j)&\equiv& h(\o_j).
\labels{csigma}\ee

The explicit decomposition formulas for $n\le 3$ are:
\be
u_1&=&u[1]=c[1],\nn\\
u_2&=&{1\over 2}\{u[12]+u[21]\}={1\over 2}\{c[1]c[2]+c[21]\},\nn\\
u_3&=&{1\over 6}\{u[123]+u[132]++u[213]u[312]+u[231]+u[321]\}\nn\\
&=&{1\over 6}\{c[1|2|3]+c[1|32]+c[21|3]+c[31|2]+c[231]+c[321]\}.
\labels{uc3}\ee

Incidentally, the decomposition formula \eq{cp} leads to the decomposition
formula \eq{dec} in the following way. Since $u_n$ in \eq{unint} is
symmetrical in all its arguments, we may replace its integration
region $R_n$ by the hyper-cube
$W_n$, provided we compensate the repitition by dividing the
integral by a factor $n!$. Subsituting the decomposition \eq{cp}
into this hyper-cube integral, and noting from \eq{cne} and \eq{csigma}
that 
\be
C_m={i^m\over m!}\int_{W_m}d^m\o\ c[\s']\labels{cc}\ee
whatever $[\s']$ is, as long as its length is $m$, then we obtain
\eq{dec} from \eq{cp}, where $m!\xi(m_1m_2\cdots m_k)$ is the number
of ways to partition the sequence $[\s']$ of length $m$, into
$k$ subsequences, the first of length $m_1$, the second
of length $m_2$, etc. For example, on the right-hand side of
 the expression $u_3$
in \eq{uc3}, the first term has a partition 
whose lengths are (111), the seond has a
partition (12), the third and the fourth have a partition (21),
and the last two have a partition (3). So the coefficients 3!$\xi(m)$
for these partitions $(m_1m_2\cdots m_k)$ are respectivey
1,1,2, and 2, agreeing with what is given in the last equation of
\eq{u3c}.

Returning to \eq{unint} and \eq{cp}, let us define
\be
C[\s']={i^m\over m!}\int_{R_m}d^m\o\ c[\s'],\labels{csi}\ee
where $R_m$ is the $m$-dimensional hyper-triangular region
of the $\o_{\s'_i}$ variables. The normalization is defined 
so that $C_m$ is equal to $C[\s']$ summed over all the $m!$
permutations of the numbers in $[\s']$.
The decomposition for the integral $U_n$
can now be obtained from \eq{unint} 
and \eq{cp} to be
\be
U_n=\sum_{[\s]\in S_n}{1\over n!}\(\prod_{i=1}^km_i!\)
C[\s]_P,\labels{deccs}\ee
where $m_i$ is the length of the $ith$ subsequence of $[\s]_P$,
and $\sum_{i=1}^km_i=n$. This is an alternative decomposition of $U_n$.
It is asymmetrical in the energies whereas the decomposition in
\eq{dec} is symmetrical.
It is this asymmetry that will be exploited for infrared factorization.

Let us now examine the infrared property of an {\it irreducible
factor} $C[\s']$. Since  the last element of $[\s']$ is the smallest
number of that subsequence,
its corresponding $\o$-variable will be the largest
in the region $R_m$, hence from
\eq{csigma} and \eq{csi} we conclude that $C[\s']$
is of order $g^m\ln \l$, where $m$ is the length of $[\s']$.
The term $C[\s]_P$ in \eq{deccs} is therefore of order $g^n\ln^k\l$,
where $k$ is the number of subsequences of $[\s]_P$. 

In particular, the leading contribution comes from $k=n$ and the
partitioned sequence $[\s]_P=[1|2|3|\cdots |n]$. The least dominant
contribution comes from sequences $[\s]$ with $\s_n=1$ and hence 
$k=1$.

The added advantage in analyzing infrared behavior by decomposition
comes from its factorization property. Since each irreducible factor
$C[\s']$ is of order $g^m\ln \l$, to order 1 it must be of the form
$g^ma[\s']\ln(\l/\l_0[\s'])$. If the constants $a$ and $\l_0$ can
be computed for every $[\s']$, then the {\it exact} infrared behavior
of $U_n$, accurate to all powers of $\ln\l$, can be obtained from \eq{deccs} 
just by multiplication and summation. 
This property will be exploited in a future study of the infrared behavior 
of QCD.

\acknowledgements
This research is supported in part by the Natural Sciences and
Engineering Research Council of Canada, and the Fonds pour la Formation de
Chercheurs et l'Aide \`a la Recherche of Qu\'ebec. 

\appendix
\section*{}
In this appendix we write down the integrals for the low-order
quantities,  and verify directly from them
 the identities discussed in the text.
Since the expressions are long, we need some shorthand
notations. We will write $\o(123)=\o_1+\o_2+\o_3
+i\e$, and similarly for other sums of $\o_i$'s and $i\e$. 
A vertical bar will be used to denote multiplication, {\it e.g.,}
$\o(123|4|56)=\o(123)\o(4)\o(56)$.
We will also
write the nested commutator $h[\o_1\o_2\cdots\o_n]$ simply as
$h[12\cdots n]$. A vertical bar in the argument of $h[\cdots]$ will
be used to indicate multiplication. Hence $h[12|3|674]=h[12]h(3)h[674]$.
We will re-define the integration variables $\o_i$, if necessary, so
that all integrals $\int d^n\o$
appearing below are taken over the hyper-triangular region
$R_n(0-\infty)$, where $n$ is the number of integration variables.
Unless otherwise specified, all integrals below are understood to be
integrated over this $R_n$.
With this notation, the  explicit expressions can be obtained from 
\eq{vert}, \eq{cne}, \eq{eoepf}, \eq{en}, \eq{enc}. 
The identities we want to verify are \eq{u3c}, \eq{u3ed}, \eq{uue},
and \eq{bub}.

The first-order expressions are
\be
U_1=C_1=\U_{11}=\C_{11}=i\int d\o{h(\o)\over\o+i\e}.\labels{u1}\ee
The second-order unbarred quantities are
\be
U_2&=&i^2\int d^2\o\left\{ {h[1|2]\over\o(12|2)}+{h[2|1]\over \o(21|1)}
\right\},\labels{u2}\\
\h C_2&=&{i^2\over 2}\int d^2\o\left\{{h[12]\over\o(12|2)}+{h[21]\over \o(21|1)}
\right\},\labels{c2}\\
\h C_1^2&=&\h\C_{11}^2={i^2\over 2}
\int d^2\o{h[1|2]+h[2|1]\over\o(1|2)}.\labels{c1s}\ee
The reason for two terms to be present in each expression
is because a square integration
region can be written as the
sum of two triangular integration regions. 
The second equation in \eq{u3c} for $U_2$
 can now be verified using the relation
$h[12]=h[1|2]-h[2|1]$ to express all commutators in terms
of products, and the trivial identity
\be
{1\over A}+{1\over B}={A+B\over AB}.\labels{trivial}
\ee

The second-order barred quantities are as follows. 
\be
\U_{12}&=&\int_{-\infty}^\infty d\o_2\D_2(\o_2)=
i^2\int_{-\infty}^\infty d\o_2\int_{-\infty}^{\o_2}d\o_1{h[12]\over
\o(12|2)}\nn\\
&=&i^2\int d^2\o{h[21]\over\o(21|1)}
=\C_{12},
\labels{u12}\\
\U_{22}&=&\int d^2\o\D_1(\o_1)\D_1(\o_2)=i^2\int d^2\o{h[1|2]\over
\o(1|2)},\labels{u22}\\
\C_{22}&=&\int d^2\o[\D_1(\o_1),\D_1(\o_2)]=i^2\int
d^2\o {h[12]\over\o(1|2)}.\labels{c22}\ee
The second equation in \eq{uue} for $U_2$ can now be verified 
directly from \eq{u2}, \eq{u12}, and \eq{u22}. 
Similarly the identity for $\U_{22}$ in \eq{u3ed} 
and the identity for $C_2$ in \eq{bub} can both be verified.

Next, we write down the third-order unbarred quantities:
\be
U_3&=&i^3\int d^3\o\Biggl\{
{h[1|2|3]\over\o(123|23|3)}+{h[2|1|3]\over\o(213|13|3)}+
{h[1|3|2]\over\o(132|32|2)}\nn\\
&&\hskip.65in{h[3|1|2]\over\o(312|12|2)}+
{h[2|3|1]\over\o(231|31|1)}+{h[3|2|1]\over\o(321|21|1)}
\Biggr\},\labels{u3}\\
{1\over 3}C_3&=&{i^3\over 3}\int d^3\o\Biggl\{
{h[123]\over\o(123|23|3)}+{h[213]\over\o(213|13|3)}+
{h[132]\over\o(132|32|2)}\nn\\
&&\hskip.65in {h[312]\over\o(312|12|2)}+
{h[231]\over\o(231|31|1)}+{h[321]\over\o(321|21|1)}
\Biggr\},\labels{c3}\\
{1\over 6}C_1C_2&=&{i^3\over 6}\int d^3\o\Biggl\{
{h[1|23]\over\o(1|23|3)}+{h[2|13]\over\o(2|13|3)}+
{h[3|12]\over\o(3|12|2)}\nn\\
&&\hskip.65in {h[1|32]\over\o(1|32|2)}+{h[2|31]\over\o(2|31|1)}+
{h[3|21]\over\o(3|21|1)}\Biggr\},\labels{c1c2}\\
{1\over 3}C_2C_1&=&{i^3\over 3}\int d^3\o\Biggl\{
{h[23|1]\over\o(1|23|3)}+{h[13|2]\over\o(2|13|3)}+
{h[12|3]\over\o(3|12|2)}\nn\\
&&\hskip.65in {h[32|1]\over\o(1|32|2)}+{h[31|2]\over\o(2|31|1)}+
{h[21|3]\over\o(3|21|1)}\Biggr\},\labels{c2c1}\\
{1\over 6}C_1^3&=&{i^3\over 6}\int d^3\o\Biggl\{
{h[1|2|3]\over\o(1|2|3)}+{h[2|1|3]\over\o(2|1|3)}+
{h[1|3|2]\over\o(1|3|2)}\nn\\
&&\hskip.65in {h[3|1|2]\over\o(3|1|2)}+
{h[2|3|1]\over\o(2|3|1)}+{h[3|2|1]\over\o(3|2|1)}
\Biggr\}.\labels{c1c}\ee
Six terms each are present in each of these equations because a
cube contains six hyper-triangles. We can now verify directly
that the last equation in \eq{u3c} is correct. For example, take
the terms proportional to $i^3h[1|2|3]$. In order for \eq{u3c}
to be correct, we need to have
\be
{1\over\o(123|23|3)}&=&{1\over 3}\Biggl\{{1\over\o(123|23|3)}-
{1\over\o(132|32|2)}-{1\over\o(312|12|2)}+{1\over\o(321|21|1)}
\Biggr\}\nn\\
&+&{1\over 6}\Biggl\{{1\over\o(1|23|3)}-{1\over\o(1|32|2)}
\Biggr\}+{1\over 3}\Biggl\{{1\over\o(3|12|2)}-{1\over\o(3|21|1)}
\Biggr\}\nn\\
&+&{1\over 6}{1\over\o(1|2|3)}.\labels{u3id}\ee
By using \eq{trivial} repeatedly, this can be verified to be true.

The third-order $\U$'s are:
\be
\U_{13}&=&\int_{-\infty}^\infty d\o_3\D_3(\o_3)=
i^3\int_{-\infty}^\infty d\o_3\int_{-\infty}^{\o_3} d\o_1d\o_2
{h[123]\over\o(123|23|3)}\nn\\
&=&i^3\int d^3\o\Biggl\{{h[231]\over\o(231|31|1)}+
{h[321]\over\o(321|21|1)}\Biggr\}=\C_{13},\labels{u13}\\
\U_{23}&=&\int_{-\infty}^\infty d\o_1\int_{-\infty}^{\o_1}d\o_2
\(\D_1(\o_1)\D_2(\o_2)+\D_2(\o_1)\D_1(\o_2)\)\nn\\
&=&i^3\int d^3\o\Biggl\{
{h[1|32]\over\o(1|32|2)}+{h[21|3]\over\o(21|1|3)}+{h[31|2]\over\o(31|1|2)}
\Biggr\},\labels{u23}\\
\U_{33}&=&\int d^3\o \D_1(\o_1)\D_1(\o_2)\D_1(\o_3)
=i^3\int d^3\o{h[1|2|3]\over\o(1|2|3)}.\labels{u33}\ee
In order for the last equation of \eq{uue} to be true,
we need to have \eq{u3} to be the sum of the three equations above.
Taking, for example, those proportional to $i^3h[1|2|3]$. It requires
\be
{1\over\o(123|23|3)}={1\over\o(321|21|1)}-{1\over\o(1|32|2)}
-{1\over\o(21|1|3)}+{1\over\o(1|2|3)},\labels{u3ub3}\ee
which is true. Other terms can be similarly verified.

The third-order $\C$'s are:
\be
\C_{23}&=&\int_{-\infty}^\infty d\o_1\int_{-\infty}^{\o_1}d\o_2
\([\D_1(\o_1),\D_2(\o_2)]+[\D_2(\o_1),\D_1(\o_2)]\)\nn\\
&=&i^3\int d^3\o\Biggl\{
{h[132]\over\o(1|32|2)}+{h[213]\over\o(31|1|2)}+{h[312]\over\o(21|1|3)}
\Biggr\},\labels{c23}\\
\C_{33}&=&\int d^3\o[\D_1(\o_1),[\D_1(\o_2),\D_1(\o_3)]]
=i^3\int d^3\o{h[123]\over\o(1|2|3)},\labels{c33}\\
\[ \C_{11}, \C_{12}\] &=&i^3\int d^3\o\Biggl\{
{h[132]\over\o(1|32|2)}+{h[321]\over\o(3|21|1)}+{h[231]\over\o(2|31|1)}
\Biggr\}.\labels{c11c12}\ee
We shall not verify explicitly the last two equations of \eq{u3ed},
but will attempt to check the last equation of \eq{bub}, relating
$C_3$ to the $\C$'s. $C_3$ is given in \eq{c3}, and the $\C$'s above. 
All of them contain the triple nested commutator
$h[ijk]$, so it is more convenient to check the identity in \eq{bub}
by comparing the coefficients of $h[ijk]$ rather than the product
$h[i|j|k]$. There are $3!=6$ such nested commutators but they are 
related by anti-symmetry and the Jacobi identity, so there are only
two independent ones, which we may take to be $h[321]$ and $h[231]$.
The others are given in terms of these two by
\be
h[312]&=&-h[321],\quad h[123]=h[321]-h[231],\nn\\
h[213]&=&-h[231],\quad h[132]=h[231]-h[321].\labels{jacobi}\ee
To verify the $C_3$ relation in \eq{bub}, let us compare the coefficients
of $i^3h[321]$ of all the terms. If the identity holds, the following
relation must be true:
\be
&&{1\over \o(123|23|3)}-{1\over \o(132|32|2)}-{1\over \o(312|12|2)}+
{1\over \o(321|21|1)}\nn\\
&=&{1\over \o(1|2|3)}-{3\over 2}\Biggl\{{1\over \o(1|32|2)}+
{1\over \o(21|1|3)}\Biggr\}
+3{1\over \o(321|21|1)}\nn\\
&+&\h\Biggl\{-{1\over \o(1|32|2)}
+{1\over \o(3|21|1)}\Biggr\}.\labels{c3id}\ee
This is so as can be verified by using \eq{trivial}.

\end{document}